# On the knowledge production function

Boris M. Dolgonosov[1]

Haifa, Israel

**Abstract**. Knowledge amount is an integral indicator of the development of society. Humanity produces knowledge in response to challenges from nature and society. Knowledge production depends on population size and human productivity. Productivity is a function of knowledge amount. The purpose of this study is to find this function and verify it on empirical material, including global demographic and information data. The productivity function is a basic element of the theory that results in the dynamic equations of knowledge production and population growth. A separate problem is the quantitative assessment of knowledge. To solve it, we consider knowledge representations in the form of patents, articles and books. Knowledge is stored in various types of devices, which together form a global informational storage. Storage capacity is increasing rapidly as digital technology advances. We compare storage capacity with the memory occupied by the forms of knowledge representation. The results obtained in this study contribute to the theory of knowledge production and related demographic dynamics and allow us to deepen our understanding of civilization development.

**Highlights**

Human productivity in creating knowledge plays a critical role in civilization development.

For a pre-information society, the constant productivity approximation is acceptable.

In an information society, productivity increases as knowledge accumulates.

The productivity function is applied to patents, articles and books as forms of knowledge.

Productivity for articles and books is a power function of their number, and for patents it is a linear function.

---

[1] E-mail: borismd31@gmail.com. Boris M. Dolgonosov, Ph.D. (1979, Moscow State University), D.Sc. (1997, Russian Academy of Sciences), Head of the Ecological modeling laboratory, Institute of Water Problems, Russian Academy of Sciences, Moscow — until 2014. Currently: Independent researcher, Haifa, Israel. The author of more than 160 scientific publications including two monographs:
Dolgonosov, B. M., 2009. Nonlinear dynamics of ecological and hydrological processes. Moscow: Editorial URSS, Librokom. (in Russian).
Dolgonosov, B. M., Gubernatorova, T.N., 2011. Mechanisms and kinetics of organic matter destruction in aquatic environments. Moscow: Editorial URSS, Krasand. (in Russian).

# 1. Introduction

The development of society is directly related to the accumulation of knowledge both in the field of science and technology, and in the cultural and humanitarian sphere. Knowledge production occurs at a rate that depends on knowledge amount at the moment and on the population creating knowledge. In particular, knowledge production controls population growth. The corresponding dynamic equations were obtained in (Dolgonosov and Naidenov, 2006; Dolgonosov, 2020). In this approach, a crucial factor is per capita productivity $w(q,t)$ in knowledge production (not in other areas of human activity), which depends on knowledge amount $q$ and time $t$. So, knowledge production $\dot{q} = dq/dt$ can be represented in general form as

$$\dot{q} = w(q,t)N + J(t) \tag{1}$$

where $N$ is population size, $J(t)$ is an external source of knowledge. We assume in (1) that the number of knowledge producers is proportional to the total population, as is usually the case in economic models (Romer, 1986, 1990; Kremer, 1993; Dong et al., 2016, Kato, 2016). In our previous studies (Dolgonosov, 2016, 2020) we looked at the problem of knowledge production by assuming that productivity is constant. This assumption has reasonable grounds for a pre-information society with its undeveloped computing capabilities. However, now, when we have an information society, the rapid progress of computer technology and artificial intelligence is leading to increased productivity, which should be reflected in the rate of knowledge accumulation and, as a consequence, in demographic dynamics. The problem is to figure out what the function $w(q,t)$ is, how justified the constant productivity approximation is, and under what conditions it can be applied. We consider the problem in this work.

Further development of the theory requires consideration of the general case where productivity depends on accumulated knowledge. This problem has also been addressed in economic models describing the relationship between technological development and population growth. Unlike technologies, knowledge is understood somewhat more broadly: it includes all the components of human culture, which undoubtedly also influence the population to a certain extent. However, economic models capture the essential features of the phenomenon. First of all, it is worth mentioning Romer's (1986, 1990) model, which was written for technologies, but we will extend it to knowledge in general. Romer's model can be presented in the form

$$\dot{q} = w(q)N_1^\lambda \tag{2}$$

with the only difference that Romer's $q$ is the sum of technologies (but it is not all knowledge), $N_1$ is the number of only those people who work in science and technology, and per capita productivity is expressed as

$$w(q) = w_0 q^\varepsilon \tag{3}$$

where $w_0$, $\lambda$ and $\varepsilon$ are parameters (everything is in our notation). Romer ends up taking $\lambda$ and $\varepsilon$ equal to 1.

Kato (2016) analyzes a model similar to (2)-(3), with the only difference that the total population $N$ is used instead of $N_1$. The author says the following about the exponent $\varepsilon$ (in the original it is designated as $\varphi$): "When $\varphi > 1$, then the growth rate of technological progress would rise rapidly with increasing level of technology. However, such situations have not been observed in developed nations through postwar periods, so Barro and Sala-i-Martin (1992) imposed the condition $\varphi \le 1$." We use this remark when constructing the productivity function.

Kremer's (1993) model can also be represented as (1). Unlike Romer's model (2), here instead of S&T personnel $N_1$, the total population $N$ is used, but the parameters $\lambda$ and $\varepsilon$ are still equal to 1. So, instead of (3) we have

$$w(q) = w_0 q \tag{4}$$

A similar model of technology development was used by Collins et al (2013) in their evolutionary theory of long-term economic growth.

Jones (1995, 1999) modified Romer's model by setting in (3) $\varepsilon < 1$, which after a series of transformations led him to the equation

$$\frac{\dot{q}}{q} = \alpha \frac{\dot{N_1}}{N_1} \tag{5}$$

where $\alpha = \lambda/(1 - \varepsilon)$. The meaning of this equation can be clarified after integrating it, which yields

$$q = k N_1^\alpha \tag{6}$$

$k$ is a constant. From (6) it follows that the technologies accumulated to date are only the output of technology producers currently working. However, the influence of previous generations, whose work also contributed to the development of technology, is not represented in this approach. The equation for $q$ must contain an integral term summing up the contribution of past generations.

The same problem was noted by Dong et al. (2016), which based on an analysis of well-known models and extensive empirical material, showed that technological growth depends not only on the current generation of people, but also on the achievements of past generations. In addition, the relationship between technology and population is nonlinear. The authors found deviations from the proportionality law $N_1 \propto N$ between the number of technology producers and total population, when considering the long-term evolution of society over millennia.

Okuducu and Aral (2017) suggested that productivity could be a constant, linear, quadratic, or exponential function of knowledge amount, and used these representations to compute various hypothetical scenarios of knowledge dynamics.

There is a difference between the approaches from the knowledge viewpoint (1) and from the economic one (2)-(4). Productivity $w(q)$ is the annual per capita knowledge product (patents, articles, books, etc.) in the first case and gross product in the second one. Knowledge is measured in information units, and technology and gross product in monetary units.

The question arises whether the information approach to demographic dynamics is divorced from reality and whether the corresponding model can be calibrated (Court and McIsaac, 2020). The answer to this question is one of the objectives of this work. Regarding the reality and prospects of this approach, we can refer to the work (Dolgonosov, 2020), which proposed a general global-scale model, including economic, environmental, demographic and information components, and which was successfully calibrated against extensive empirical data.

In connection with the development of artificial intelligence, a dilemma has arisen about how to describe the presence of intelligent machines, whether to include them among the producers of knowledge, thereby expanding the number $N$, or to continue to believe that knowledge is produced by people, and the machine is for now just a tool that helps them in the production of knowledge. Sadovnichy, Akaev and Korotayev (2022) develop the former approach, believing that intelligent machines can now be considered producers of knowledge and hence included in the number $N$ along with humans. This is a promising direction, especially given the rapid development of AI. But for now, following the analysis of Akaev and Sadovnichy (2021), we will remain with the traditional approach, according to which it is people who produce knowledge, and intelligent machines only help them in this matter. Then the effect of AI manifests itself through an increase in the amount of knowledge and a corresponding increase in human productivity.

The above-mentioned productivity functions proposed by various authors require verification based on empirical material. To this end, we revisit the issue of productivity as a function of knowledge and verify the theoretical results using literature data.

Another nontrivial problem is how to determine the amount of knowledge. The most consistent approach is to estimate memory capacity the knowledge takes up. However, at the moment such information is unlikely to exist. Meanwhile, there is evidence that digital memory is rapidly increasing over time, in what appears to be a global information explosion during the digitization period (1986-2007 onwards) (Hilbert, 2014).

It should be expected that the total memory capacity far exceeds knowledge capacity due to the multiple replication of useful information, especially in graphic and video formats. In this situation, it is necessary to use data on different types of knowledge representation, such as patent applications, original articles and books. These data have been largely cleared of duplication. Knowledge production should be assessed separately for each type. Below we use this approach.

## 2. Information and knowledge

Information is signals received from the world and stored in memory (Chernavsky, 2004). Knowledge is a meaningful piece of information in the form of models of the world. Models contain information about objects (declarative knowledge) and algorithms of processes (procedural knowledge) (Burgin, 2017). The concepts of declarative and procedural knowledge were introduced by ten Berge and van Hezewijk (1999), although the rationale of this classification focused on the psychological aspects and characteristics of human memory. Here we will look at these concepts for various types of memory.

Memory can be varied in nature. In relation to a person, it can be internal and external. Human internal memory is represented predominantly by neuronal and genetic types. Information in internal memory accumulates and remains throughout a person's life. Some of this information is transmitted to other people, some is recorded in external storage, and the rest is lost.

External memory includes a variety of digital and analog storage devices and is usually much more durable (except perhaps for genetic memory). To reliably store knowledge, it is duplicated in memory. The higher the value of this knowledge, the higher the frequency of duplication. Due to the need for such duplication, as well as for storing a large amount of information that has not yet been processed and comprehended (which is not yet knowledge), the storage capacity must significantly exceed the amount of knowledge. Informational storage refers to all installed devices, the capacity of which determines the maximum available memory. The amount of information in it is measured in optimally compressed bytes (Hilbert, 2014).

### 3. Model

#### *3.1. Knowledge production and accumulation*

The need to solve non-standard problems that life poses to people encourages knowledge production (Fig. 1). Knowledge is professionally produced only by part of the population. As in many economic models, we assume that this part is proportional to population size. Dong et al. (2016) found deviations from this law for individual countries, but there are reasons to believe that the deviations are likely to be smoothed out when moving to a global scale as usually happens when a statistical system is enlarged. Then the overall rate of knowledge production will be equal to average productivity multiplied by population size, which is generally expressed by equation (1). However, for humanity as a global system, this equation can be simplified by keeping in mind the following. Human civilization does not have extraterrestrial contacts, hence there are no external sources of knowledge, so in (1) we must put $J = 0$. Due to this isolation, the system is autonomous, which means that productivity $w$ does not depend on time explicitly, but only through $q(t)$. As a result, equation (1) is reduced to the form

$$\dot{q} = w(q)N \qquad (7)$$

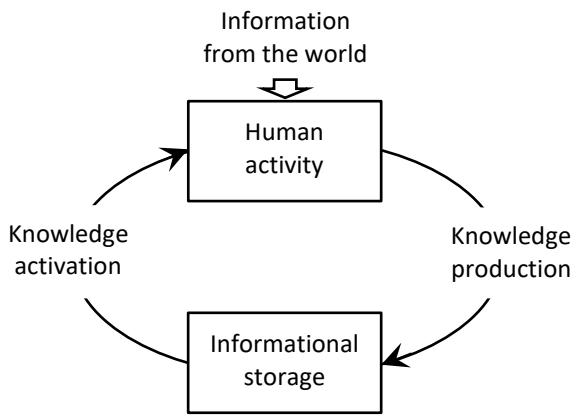

Fig. 1. Conceptual diagram of knowledge production and accumulation.

Equation (7) can be written as

$$\frac{dq}{w(q)} = N(t)dt \qquad (8)$$

Integrating (8) with the initial condition

$$t = t_0, \quad q = q_0 \tag{9}$$

and introducing functions

$$F(q) = \int_{q_0}^{q} \frac{dq'}{w(q')} \tag{10}$$

$$S(t) = \int_{t_0}^{t} N(t')dt' \tag{11}$$

we come to the equation

$$F(q) = S(t) \tag{12}$$

which implicitly specifies $q$ as a function of the cumulative population number $S(t)$, thereby formalizing knowledge accumulation over time.

### 3.2. Productivity function

Based on general considerations, productivity as a function of knowledge should have the following properties:

- in the limit of zero knowledge ($q = 0$), productivity has a nonzero value $w(0) = w_0$ (at $w_0 = 0$ knowledge production will never begin);
- at a high level of knowledge, productivity increases slowly according to the power law $w(q) \propto q^\varepsilon$ with an exponent $\varepsilon$ *not exceeding* 1 (because an average knowledge producer uses a limited amount of knowledge in his creative process — this is consistent with Barro and Sala-i-Martin's (1992) point of view mentioned in Introduction).

The simplest interpolation formula with these properties is

$$w(q) = w_0(1 + hq)^\varepsilon \tag{13}$$

$$w_0 > 0, \quad h \geq 0, \quad 0 \leq \varepsilon \leq 1 \tag{14}$$

where $w_0, h, \varepsilon$ are parameters. If $hq \ll 1$, we can use the constant productivity approximation as in our previous works. Substitution of (13) into (10) yields

$$F(q) = \frac{1}{hw_0}\big(\ln_\varepsilon(1 + hq) - \ln_\varepsilon(1 + hq_0)\big) \tag{15}$$

and according to (12) we find

$$1 + hq = (1 + hq_0) \exp_\varepsilon \left( \frac{hw_0 S}{(1 + hq_0)^{1-\varepsilon}} \right) \qquad (16)$$

where we use the deformed logarithm and the deformed exponential, which are defined as (Umarov et al., 2008)

$$\ln_\varepsilon(x) = \frac{x^{1-\varepsilon} - 1}{1 - \varepsilon} \qquad (17)$$

$$\exp_\varepsilon(x) = (1 + (1 - \varepsilon)x)^{1/(1-\varepsilon)} \qquad (18)$$

In the limit $\varepsilon \to 1$, we get the natural logarithm and exponential:

$$\ln_1(x) = \ln(x), \quad \exp_1(x) = \exp(x) \qquad (19)$$

At the ends of the $\varepsilon$ range, we have:

- a constant productivity

$$\varepsilon = 0, \; w = w_0, \; q = q_0 + w_0 S \qquad (20)$$

- productivity as a linear function of knowledge

$$\varepsilon = 1, \; w = w_0(1 + hq), \; 1 + hq = (1 + hq_0) \exp(hw_0 S) \qquad (21)$$

In (20) and (21), accumulated knowledge is, respectively, a linear and exponential function of the total number of people $S$ over the period under study $(t_0, t)$. All people are taken into account here, not just the direct producers of knowledge, since the number of producers is assumed to be proportional to the population.

The presence of the integral quantity $S$ in (16) describes the contribution of past generations to the accumulation of knowledge, as discussed by Dong et al. (2016), in contrast to formula (6), which refers only to the current population.

### 3.3. Asymptotics

Let us consider a situation where the most probable values of the parameters in equation (16) correspond to the limit $h \to \infty$. Minimizing the standard deviation of the model from data by varying $h$ causes $w_0$ to depend on $h$. The asymptotic form of equation (16) is

$$q \approx (q_0{}^{1-\varepsilon} + (1 - \varepsilon)h^\varepsilon w_0 S)^{1/(1-\varepsilon)} \qquad (22)$$

In the limit $h \to \infty$, expression (22) must be independent of $h$, which implies

$$w_0 \approx ch^{-\varepsilon} \tag{23}$$

and

$$q \approx q_0 \left(1 + \frac{(1-\varepsilon)cS}{q_0^{1-\varepsilon}}\right)^{1/(1-\varepsilon)} \tag{24}$$

where $c$ is a positive constant. Productivity (13) asymptotically obeys the power law

$$w(q) \approx cq^{\varepsilon} \tag{25}$$

Thus, the general productivity function (13) includes three special cases: a constant (20), linear (21) and power (25) function. There is another special case, which we consider in the next item.

### 3.4. Exponential productivity

Okuducu and Aral (2017) considered productivity as an exponential function of $q$ as one of the options. This option leads to a singularity, but for the sake of completeness we will consider it. Formula (13) covers this option if the coefficient $h$ decreases with increasing $\varepsilon$ according to the law $h = a\varepsilon^{-1}, \ a > 0$, and if, unlike (14), there is no upper limit for $\varepsilon$. Then $w(q) = w_0(1 + a\varepsilon^{-1}q)^{\varepsilon}$, and in the limit $\varepsilon \to \infty$ we get

$$w(q) = w_0 e^{aq} \tag{26}$$

From (10)-(12) it is easy to find

$$q = q_0 + \frac{1}{a}\ln\frac{1}{1 - bS} \tag{27}$$

where $b = aw_0 e^{aq_0}$. Since the cumulative population number $S(t)$ increases with time, at some point in time it reaches the value $S = 1/b$, at which a singularity occurs. Thus, in a finite time the accumulated knowledge $q$ becomes infinite, which is physically impossible.

# 4. Model calibration

## 4.1. From continuous to discrete

The productivity function (13) is calibrated by varying its parameters in order to minimize the standard deviation from data. Due to the annual discreteness of demographic data, integral (11) should be replaced by the sum of population over years $t_0$ to $t$:

$$S = \sum_{i=t_0}^{t} N_i \tag{28}$$

where $N_i$ is the $i$th year population, $i$ is a year number.

Knowledge can be represented in different forms, of which we will consider three: patents, articles, and books. Each form of knowledge, accumulated up to a certain year $t$ inclusive, represents the sum

$$q = q_0 + \sum_{i=t_0}^{t} X_i \tag{29}$$

where $X$ is knowledge production measured on a case-by-case basis by the annual publication of patents, articles or books (what is denoted as $\dot{q}$ in the basic equation (7)), $X_i$ corresponds to $i$th year, $q_0$ is knowledge (number of patents, articles, or books) accumulated up to year $t_0$ (not including $t_0$ itself). This equality is also used to determine informational storage capacity.

## 4.2. Data

To calibrate the model equations (16) and (24), we used the literature data presented in Fig. 2. Articles in scientific and technical journals and patent applications are represented by global data (WB, 2022; OECD, 2022), articles for 2000-2018, patents for 1985-2020. New book title data is selected for a group of 30 countries based on information provided by Fink-Jensen (2015). The group composition is indicated in the note to Table 1. The criterion for including a particular country in the group is the availability of data on books published for 1950-1996. For other countries, the data range is less than specified. There are gaps in the data for individual years, which are filled by linear interpolation. When calibrating the model, we used the group population for books, and the world population for articles and patents (Fig. 3).

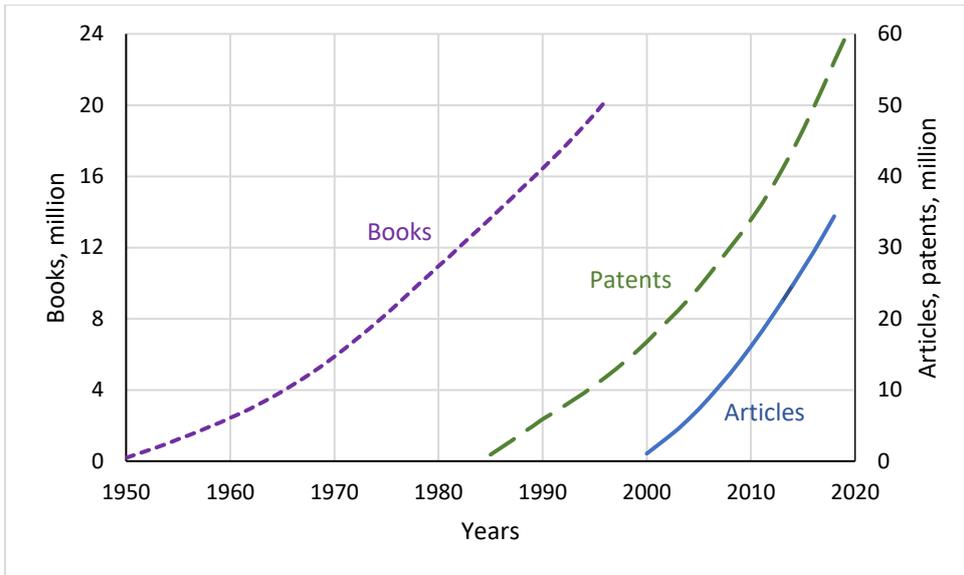

Fig. 2. Cumulative sums of patents, articles and books over the years of observation. Patents and articles represent global data, while books refer to the group of 30 countries listed in the note to Table 1. Data sources: number of scientific and technical journal articles — WB, 2022; number of patent applications — OECD, 2022; number of new book titles — Fink-Jensen, 2015.

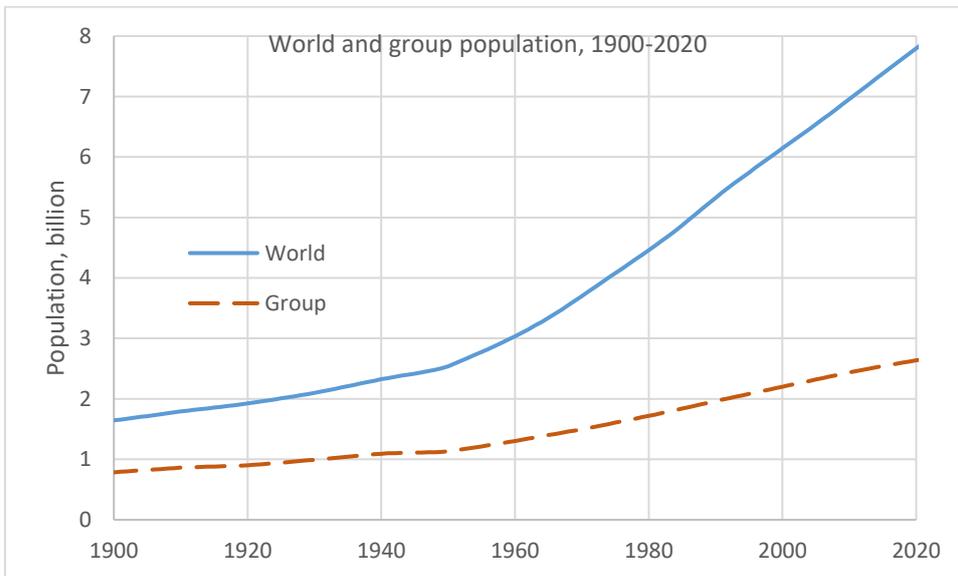

Fig. 3. Population of the world and the group of 30 countries over time. See note to Table 1 for the group composition. Data sources: UN, 2022; Gapminder, 2022.

### 4.3. Initial amount of knowledge

The informational storage capacity at the onset of the digitization period is known from the literature (Hilbert, 2014). However, this cannot be said about the initial amount of knowledge $q_0$, represented in the form of patents, articles, and books. To find $q_0$, we use an indirect estimate based on the relationships between annual knowledge production $X(t)$ (as denoted in (29)), gross domestic product $G(t)$, and population $N(t)$. All these quantities are provided by literature data (for references, see the captions to Fig. 2 and Fig. 3). The problem is that the time series $X(t)$ is usually very short, and in order to find $q_0$ it is necessary to sum $X(t)$ over a fairly long retrospective period. This can be done using the following algorithm:

1° generate a function $N(t)$ based on demographic data;

2° generate functions $X(G)$ and $G(N)$ on ranges provided by data;

3° approximate $X(G)$ and $G(N)$ with suitable functions and continue the functions to the origin (where $G$ and $N$ are zero);

4° make up a composition of functions $X(t) = X\left(G\big(N(t)\big)\right)$, continuing it into the distant past, where $X$ tends to zero;

5° take the sum of $X(t)$ for the entire previous period up to point $t_0$ (not including it), where the data for $X$ begins:

$$q_0 = \sum_{i=-\infty}^{t_0-1} X_i \qquad (30)$$

Formally, the summation starts from $-\infty$, but in fact it is acceptable to take a fairly distant point in the past, where $X(t)$ is very small. Such a point here is 1900, when the production of patents, articles and books is insignificant compared to modern amounts.

6° calculate $q(t)$ using formula (29) in two ways: (i) using the available data for $X$, and (ii) using the results of model calculations according to item 4° (to compare the model with the data).

An example of applying this algorithm to finding the initial number of articles $q_0$ accumulated by the year $t_0 = 2000$ is shown in Fig. 4. Data on articles are available in the range 2000-2018. Despite such a short data range, the use of this algorithm makes it possible to estimate the accumulation of articles in a much wider range: 1900-2020. The agreement between the model and the data is satisfactory. This algorithm was also applied to patents and books (Fig. 5).

## 5. Results and discussion

The parameter values found as a result of model calibration are presented in Table 1 and Fig. 6. The accuracy of matching the model with the data is very high, as evidenced by the determination coefficient $R^2$, the values of which are close to 1.

### 5.1. Storage capacity

The best fit of equation (16) to the data is achieved at $\varepsilon = 1$, when a linear productivity (21) is the case:

$$q = q_h \left( \rho e^{S/\sigma} - 1 \right) \tag{31}$$

$$q_h = 2.053, \ \ \rho = 2.266, \ \ \sigma = 29.42 \tag{32}$$

where

$$q_h = \frac{1}{h}, \ \ \rho = 1 + hq_0, \ \ \sigma = \frac{1}{hw_0} \tag{33}$$

$q$ is measured in Exabytes (only in this case), $S$ and $\sigma$ are measured in billion people×year.

### 5.2. Patents

The number of patents is also best suited to the linear case $\varepsilon = 1$, see (21), and obeys equation (31) with parameters (33) having values

$$q_h = 20.41, \ \ \rho = 1.770, \ \ \sigma = 230.8 \tag{34}$$

here and further in (35) $q$ is measured in million texts.

### 5.3. Articles

Equation (16) when applied to the number of scientific and technical journal articles gives the best result in the asymptotic limit $h \to \infty$, which corresponds to equation (24) at $\varepsilon = 0.7580$ (Table 1). Equation (24) can be rewritten as

$$q = q_0 \left( 1 + \frac{S}{\sigma \tau} \right)^{\tau} \tag{35}$$

$$q_0 = 20.04, \ \ \sigma = 114.5, \ \ \tau = 4.132 \tag{36}$$

where

$$\sigma = \frac{q_0^{1-\varepsilon}}{c}, \ \ \tau = \frac{1}{1-\varepsilon} \tag{37}$$

### 5.4. Books

For the number of new book titles (in all kinds of literature), the best result corresponds to the same asymptotic formula (35) as for articles, with $\varepsilon = 0.5814$ and parameter values

$$q_0 = 8.749, \ \ \sigma = 46.74, \ \ \tau = 2.389 \tag{38}$$

### 5.5. Memory capacity assessment

To estimate the memory capacity (in bytes) occupied by patents, articles and books, we use estimates of the average sizes of the mentioned texts. Analysis of samples of several hundred patents and articles gives an average size of approximately 1.5 Megabytes per patent (or article). Similarly for books, we get an average size of 14 Megabytes per book. The latest value of storage capacity (310 Exabytes) dates back to 2007. Estimates of memory amount for different types of knowledge representation as of 2007 are shown in Table 2.

We can see that the amount of memory occupied by each type of text is 6 orders of magnitude less than the total storage capacity. Storage capacity is filled mainly with visual information (photos, films, archives of TV programs, video monitoring, digitized museum exhibits, etc.). It is also necessary to consider the repeated duplication of visual and textual information, copied by almost every interested user to their devices. The need to store such immense information causes an accelerated growth in the capacity of storage devices, which is what we are seeing in reality (Fig. 6a).

Table 1. Optimal parameter values of the productivity function (13) and its asymptotics (25) for storage capacity and various types of knowledge representation*

| Model parameters | Storage 1986-2007 | Patents 1985-2020 | Articles 2000-2018 | Books 1950-1996 |
|---|---|---|---|---|
| $q_0$ | 2.6 | 15.70 | 20.04 | 8.75 |
| $\varepsilon$ | 1 | 1 | 0.7580 | 0.5814 |
| $h$ | 0.487 | 0.0490 | — | — |
| $w_0$ | 0.06978 | 0.08841 | — | — |
| $c$ | — | — | 0.01804 | 0.05304 |
| $R^2$ | 0.9963 | 0.9991 | 0.9997 | 0.9977 |

* Notes:

1) The storage capacity and the number of texts (patents, articles, or books) accumulated by the

beginning of the proper observation period are designated as $q_0$.

2) System of units: $q$, Exabytes (Exa = $10^{18}$) for storage capacity; $q$, million texts for patents, articles and books; $N$, billion people; $t$, year.

3) The determination coefficient $R^2$ for articles and books is highest for the asymptotic formula (24).

4) Data on books are given for a group of 30 countries for which data are available over the entire specified period 1950-1996 (gaps for individual years are filled by linear interpolation). The group includes countries: Argentina, Australia, Austria, Belgium, Bulgaria, Denmark, Estonia, Finland, France, Germany, Greece, Hungary, Iceland, India, Italy, Japan, Latvia, Lithuania, Netherlands, Norway, Poland, Portugal, Romania, Russian Federation, Spain, Sweden, Switzerland, Turkey, United Kingdom, United States.

Table 2. Memory capacity of informational storage and different types of knowledge representation as of 2007

| Type | Number of texts (in 2007), million | Specific capacity, Megabyte per text | Total capacity, Petabyte* |
|---|---|---|---|
| Storage (world) | — | — | 310 000 |
| Patents (world) | 44.0 | 1.5 | 0.07 |
| Articles (world) | 30.6 | 1.5 | 0.05 |
| Books (group) | 30.4 | 14 | 0.30 |

* 1 Petabyte = $10^{15}$ bytes

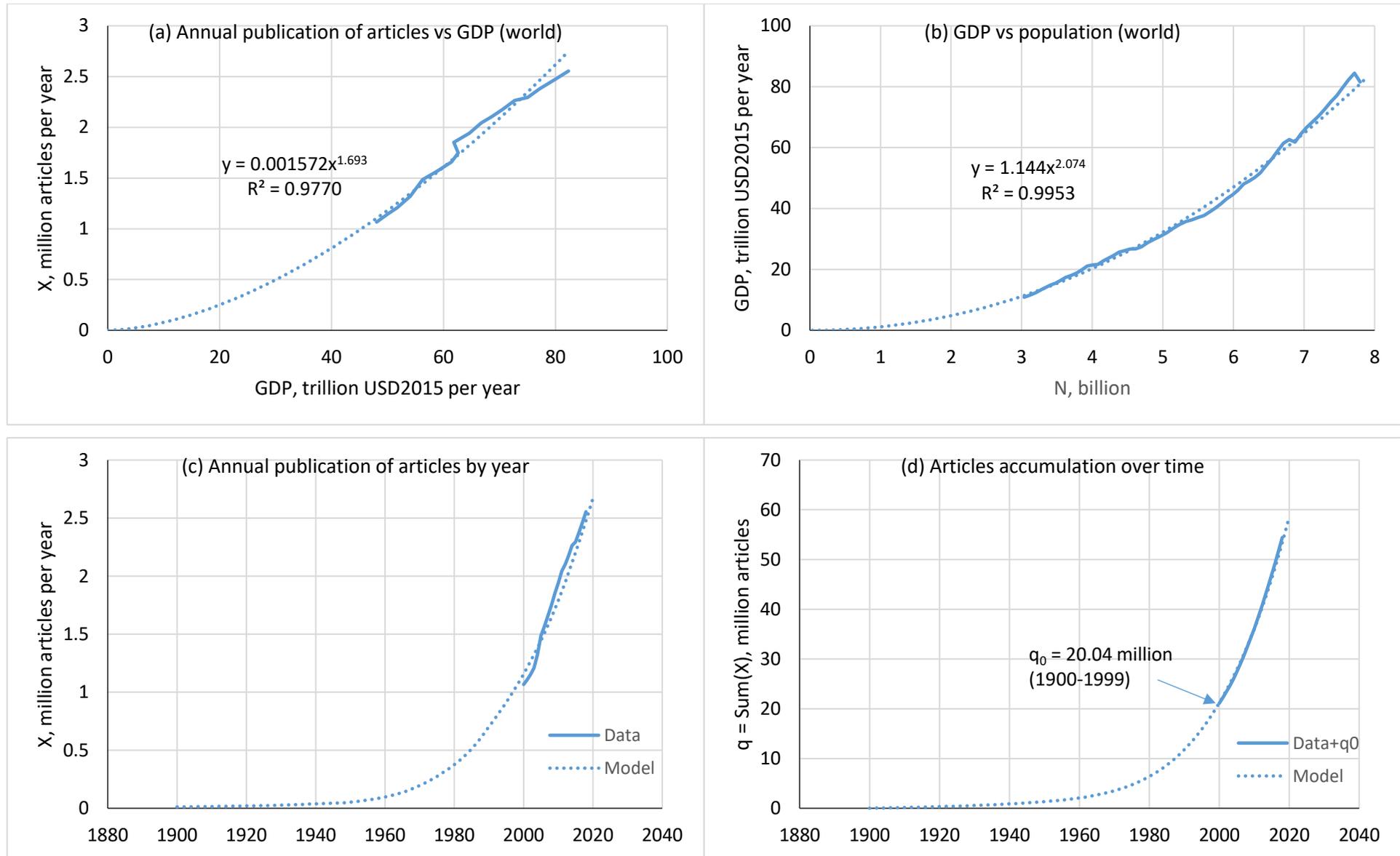

Fig. 4. Finding the number of articles $q_0$ accumulated over previous years (1900-1999) by the beginning of the observation period (2000-2018): (a) annual publication of articles vs GDP: $X(G)$; (b) GDP vs population: $G(N)$ (population $N(t)$ over time is shown in Fig. 3); (c) annual publication of articles over time: $X(t) = X\left(G\big(N(t)\big)\right)$; and finally (d) articles accumulation over time (from 1900): $q = \text{Sum}\big(X(t)\big)$. Model calculations are compared with the data.

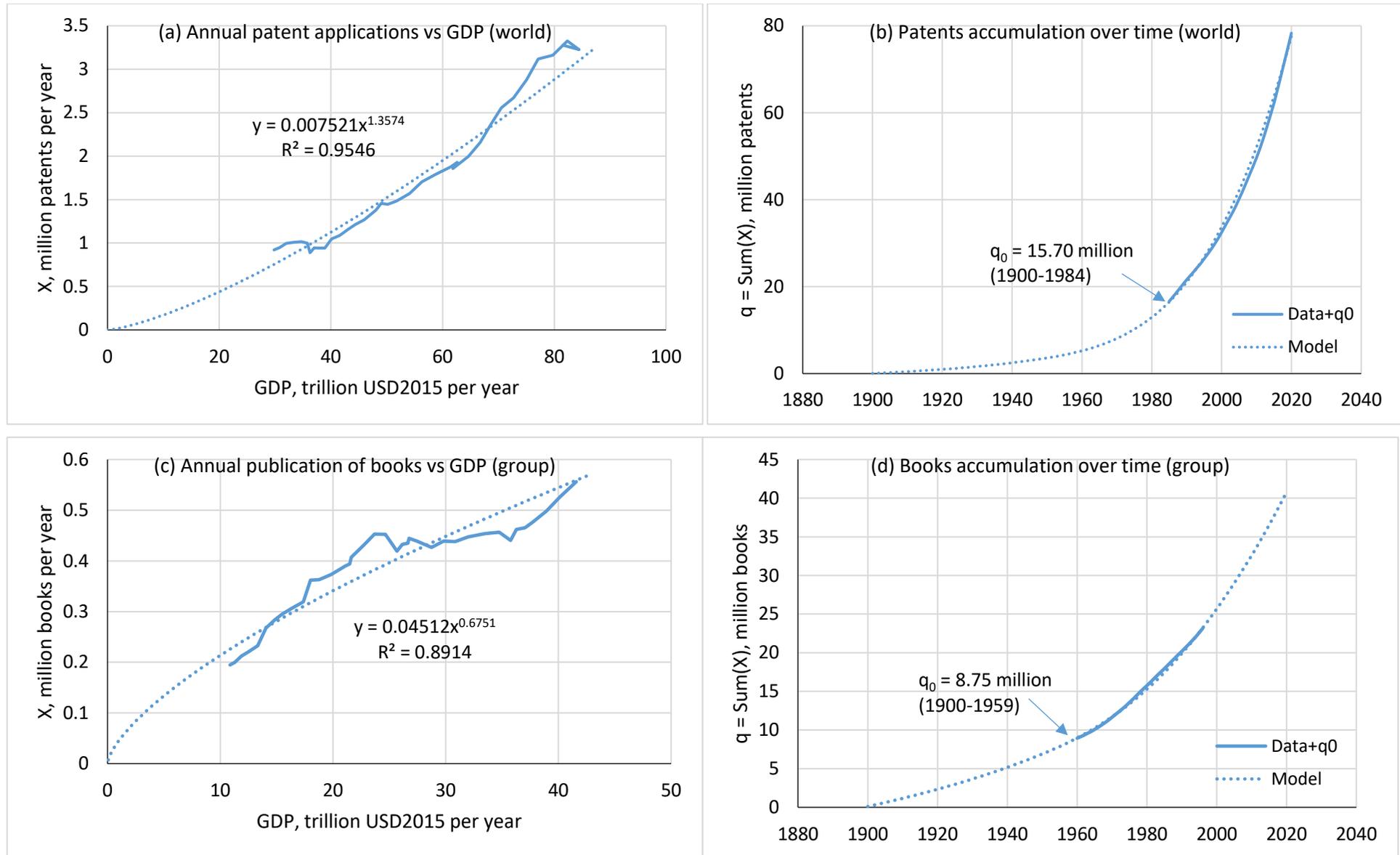

Fig. 5. Finding the number $q_0$ of accumulated patents (a, b) and books (c, d). Here, in contrast to Fig. 4, only the initial and final charts are shown.

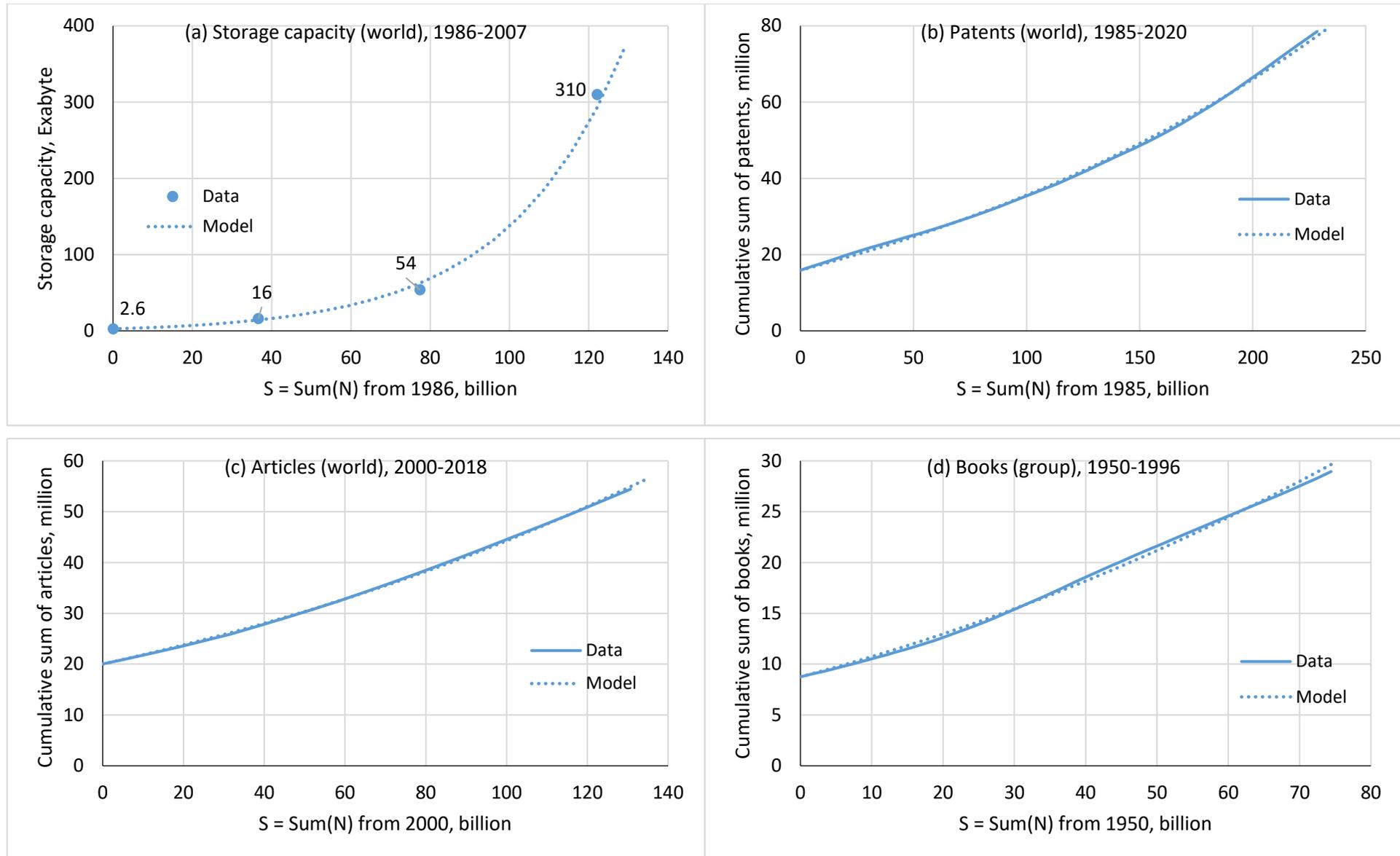

Fig. 6. (a) Storage capacity (four points correspond to 1986, 1993, 2000 and 2007), and cumulative sums of (b) patents, (c) articles, and (d) books versus cumulative sum of population during the proper observation period indicated at the top of the panels. Markers and solid lines are data, dotted lines are model. See Table 1 for model parameters. Data source for storage capacity: Hilbert (2014). Data sources for patents, articles, books, and population are indicated in the captions to Fig. 2 and Fig. 3.

*5.6. Productivity increase*

According to the adopted model, productivity increases for all types of texts studied here (patents, articles and books), as depicted in Fig. 7. When knowledge amount increases by 5 times ($q$ is from 10 to 50 units), productivity increases by 2.3, 2.5 and 3.4 times for patents, books and articles, respectively. For the same increase in storage capacity, productivity increases by 4.3 times. So, productivity grows slower than knowledge.

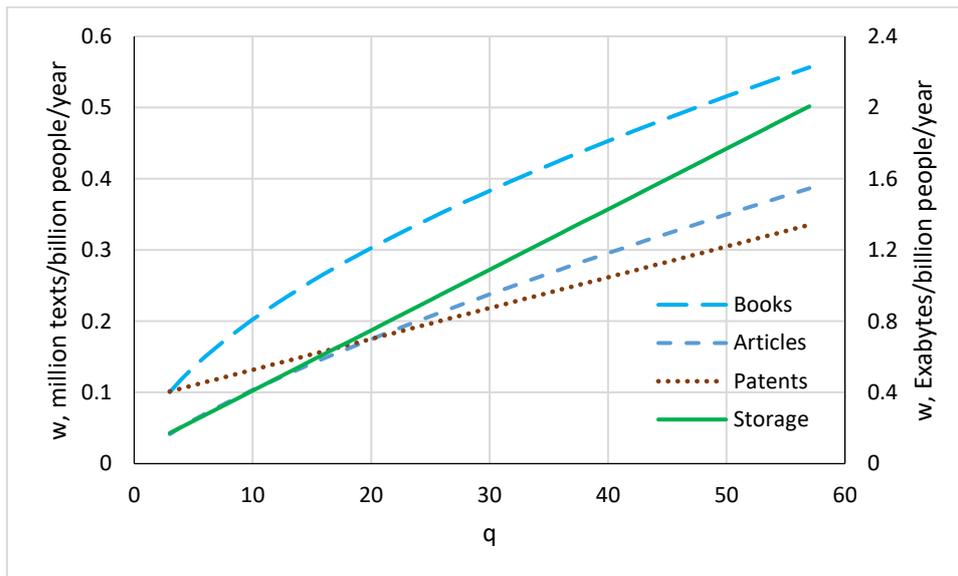

Fig. 7. Productivity $w$ as a function of knowledge amount $q$ for patents, articles and books (their $w$ is on the left axis, and $q$ is measured in million texts) and for storage capacity (its $w$ is on the right axis, and $q$ is measured in Exabytes).

Table 3 shows that during the observation period productivity increases by 2 – 2.7 times. Unlike knowledge, the informational storage stands apart: its capacity $q$ increased 113 times during the observation period, and its productivity $w$ increased 63 times. We can see that memory expands much faster than new texts (patents, articles, books) are created. Apparently, producing storage devices is a simpler process than creating new knowledge.

Table 3. The increase in productivity over the observation period

| Type | Year | $q$ | $w$ | $w_2/w_1$ |
|---|---|---|---|---|
| Storage* | 1986 | 2.6 | 0.1581 | 63.4 |
| | 2007 | 292.8 | 10.02 | |
| Patents** | 1985 | 15.92 | 0.1574 | 2.69 |
| | 2020 | 78.51 | 0.4234 | |
| Articles** | 2000 | 20.04 | 0.1750 | 2.15 |
| | 2018 | 54.42 | 0.3757 | |
| Books** | 1950 | 8.749 | 0.1872 | 2.03 |
| | 1996 | 28.94 | 0.3805 | |

*For storage: $q$, Exabytes; $w$, Exabytes per billion people per year.

**For patents, articles, and books: $q$, million texts; $w$, million texts per billion people per year.

### 5.7. Constant productivity approximation

Consider the condition under which the constant productivity approximation may be acceptable. According to (13), this condition is $q \ll q_h$, where $q_h = 1/h$ is a threshold value. Referring to Table 1, we find $q_h = 2.053$ for storage and $q_h = 20.41$ for patents. The former corresponds to 1983, the latter to 1989.

For articles and books, their productivity $w$ and accumulated knowledge $q$ obey nonlinear laws (25) and (35). As shown above, see (20), constant productivity induces a linear increase in knowledge. Equation (35) can be linearized if the condition $S \ll \sigma$ is satisfied, then $w \approx c q_0^\varepsilon$. According to (36) and (38), $\sigma = 114.5$ for articles and $\sigma = 46.74$ for books. The threshold value $S = \sigma$ is reached in 2016 for articles and in 1982 for books.

So, we can use the constant productivity approximation (20) as long as we do not get too close to the specified dates, staying in the range of $q$ where the condition $q \ll q_h$ for storage and patents or $S \ll \sigma$ for articles and books holds. To summarize, as we approach the 1980s, the constant productivity approximation loses its adequacy (for articles it happens later).

The dependence of knowledge production on population size (7), supplemented by the equation of knowledge dynamics, allows us to obtain the equation of demographic dynamics (Dolgonosov, 2016). The constant productivity approximation $w = w_0$ leads to the well-known hyperbolic law of world population growth (von Foerster et al., 1960), which operated for over a thousand years. However, deviations from this law become increasingly apparent as we approach the 1980s, which is associated with a significant accumulation of knowledge and an increase in productivity — it can no longer be considered constant. This fact is usually considered as a demographic and technological phase transition (Korotayev et al., 2015; Grinin et al., 2020a,b), and at the same time it can be interpreted as a transition from a *pre-information society*, where the constant productivity approximation operates, to a more developed *information society* with advanced computer technologies and growing human productivity.

After 1980s, personal computers became widespread and the information society continued to develop. Digital memory grew, reaching the level of analog memory and then surpassing it. The share of digital memory increased as follows: 0.8% in 1986, 3% in 1993, 25% in 2000, 94% in 2007 (Hilbert and López, 2011). The capacities of both types of memory became equal in 2003. Thus, the early 2000s can be considered a milestone in the maturation of digital civilization. Currently, the majority of world's technological memory is organized in the most accessible and fastest digital format.

## 6. Conclusion

Knowledge amount correlates with the number of patents, articles and books published in the world over the entire previous period, which makes it possible to track the dynamics of knowledge accumulation. The production of knowledge depends on its current amount and population size. The problem is to find out the form of this dependence and check how well it corresponds to real data. This dependence plays a crucial role in knowledge dynamics and related demographic dynamics.

We proposed a model in which the total rate of knowledge production is expressed as the product of human productivity and population size. Productivity increases as knowledge accumulates and information technology advances. At an early stage of societal development, knowledge is very scarce (which is typical for an underdeveloped society) and productivity is minimal, but still different from zero, ensuring the development of society. As knowledge grows, productivity gradually increases, reaching high values in a developed information society. In the asymptotic limit, when knowledge amount $q$ becomes large, the productivity behavior can be described by a power-law dependence on $q$. To combine the extreme cases of an underdeveloped and a highly developed society, we describe productivity by the interpolation dependence, which is a linear form of $q$ raised to a certain power. This dependence generalizes important special cases where productivity reduces to a constant, linear, power, or exponential function of knowledge amount.

In a developed society, information is stored primarily in digital format on various types of devices, which together form an informational storage. With the development of digital technology, storage capacity is rapidly increasing. To describe this process, we used the constructed model.

The model is calibrated using literature data for the world as a whole (applied to patents, articles and informational storage) and for the group of 30 countries (applied to books, given the lack of data for many countries) and shows good agreement with the data. It is shown that the general dependence of human productivity on knowledge amount comes down to the special cases: a linear function of $q$ for patents and storage capacity, and a power function of $q$ for articles and books. In a pre-information society, one can take advantage of the constant productivity approximation due to the relatively small amount of knowledge. A developed information society is characterized by a significant predominance of digital memory over analogue one. The population's need for repeated duplication of useful information leads to a rapid increase in the number of storage devices and hence an increase in the total capacity of informational storage, which by 2007 exceeded the memory capacity for patents, articles and books by 6 orders of magnitude. The results obtained open up an opportunity to advance in describing the dynamics of civilization development.

**Conflicts of interest:** The author declares no conflict of interest.